\documentclass[twocolumn,showpacs,preprintnumbers,amsmath,amssymb,aps,floatfix,eps,superscriptaddress]{revtex4}

\usepackage{graphics}	% Include figure files
\usepackage{bm}		% bold math

\begin{document}

\title{Dynamics of semi-flexible polymer solutions in the highly entangled regime}

\author{Manlio Tassieri\footnote[2]{Electronic address: \texttt{m.tassieri@leeds.ac.uk}}}
\affiliation{School of Physics and Astronomy, University of Leeds, LS2 9JT, U.K.}

\author{R. M. L. Evans}
\affiliation{School of Physics and Astronomy, University of Leeds, LS2 9JT, U.K.}

\author{Lucian Barbu-Tudoran}
\affiliation{Institute of Molecular and Cellular Biology, Faculty of Biological Sciences, University of Leeds, LS2 9JT, U.K.}
\author{Nasir Khan}
\affiliation{Institute of Molecular and Cellular Biology, Faculty of Biological Sciences, University of Leeds, LS2 9JT, U.K.}
\author{John Trinick}
\affiliation{Institute of Molecular and Cellular Biology, Faculty of Biological Sciences, University of Leeds, LS2 9JT, U.K.}

\author{Tom A. Waigh\footnote[1]{Electronic address: \texttt{thomas.waigh@manchester.ac.uk}}}
\affiliation{Biological Physics, School of Physics and Astronomy, University of Manchester, M60 1QD, U.K.}

\date{August 3, 2007}

\begin{abstract}
We present experimental evidence that the effective medium approximation (EMA), developed by D.C. Morse [Phys.\ Rev.\ E~{\bf 63}, 031502, (2001)], provides the correct scaling law of the macroscopic plateau modulus $G^{0}\propto\rho^{4/3}L^{-1/3}_{p}$ (where $\rho$ is the contour length per unit volume and $L_{p}$ is the persistence length) of semi-flexible polymer solutions, in the highly entangled concentration regime. Competing theories, including a self-consistent binary collision approximation (BCA), have instead predicted $G^{0}\propto\rho^{7/5}L^{-1/5}_{p}$. We have tested both the EMA and BCA scaling predictions using actin filament (F-actin) solutions which permit experimental control of $L_p$ independently of other parameters. A combination of passive video particle tracking microrheology and dynamic light scattering yields independent measurements of the elastic modulus $G$ and $L_{p}$ respectively. Thus we can distinguish between the two proposed laws, in contrast to previous experimental studies, which focus on the (less discriminating) concentration functionality of $G$.
\end{abstract}


\maketitle

Despite their importance to soft-matter physics, biology and industrial processing, the viscoelastic properties of semi-flexible polymer solutions are still not well understood and a basic analytical model has not yet been agreed upon. All current models describing the viscoelastic properties of semi-flexible polymer solutions are elaborations on the early models of Doi and Edwards \cite{DoiEdwards1,DoiEdwards2}. They developed two full theories of the entangled state for two extreme cases: completely flexible \cite{DoiEdwards1} and rigid-rod \cite{DoiEdwards2} polymers. Solutions of semi-flexible polymer, that lie between those extremes, have many regimes of viscoelastic behavior (requiring many theoretical models \cite{Mason, Odijk, Semenov, Käs, MacKintosh, Isambert, Morse1, Morse2, Morse3, Morse4}), depending on the polymers' degree of rigidity (described in terms of persistence length $L_{p}$), on their contour length $L$, and on the concentration (from dilute to highly entangled regimes). We shall focus on highly entangled isotropic solutions of semi-flexible polymers, with $L/L_{p}\sim1$. 
In particular, we study the range of concentration ($\approx$ 0.1--1 mg/mL in this case) where the geometrical mesh size $L_{m}$ is much less than $L_{p}$, and the tube diameter and entanglement length are also expected to be much less than $L_{p}$. This range of concentration was defined by Morse \cite{Morse1} as the \emph{tightly entangled regime}, and is particularly relevant to many biological and industrial polymeric fluids. In order to describe the viscoelastic behavior of the polymer network in this range of concentrations, Morse developed two analytical approximations describing the confinement forces acting on a randomly chosen test chain embedded in a ``thicket" of uncrossable chains: the binary collision approximation (BCA) and effective medium approximation (EMA). In fact, the scaling relation resulting from the BCA had previously been obtained by several others authors \cite{Semenov, Käs, MacKintosh, Isambert}, but Morse has also estimated the prefactors. So, prior to the introduction of the EMA, there was broad agreement regarding the scaling law. The approximations are summarized as follows.

(i) \textbf{The binary collision approximation} gives a rather detailed description of the interaction of a test chain with individual nearby medium chains, but neglects any effects arising from the collective elastic relaxation of the network. It yields the following expression for the elastic modulus:
\begin{equation}
\label{BCAG}
	G\approx0.40k_{B}T\rho^{7/5}L^{-1/5}_{p}.
\end{equation}

(ii) \textbf{The effective medium approximation} starts from a very different point of view, by treating the network surrounding the test chain as an elastic continuum with a shear modulus equal to the self-consistently determined plateau modulus of the solution, and the test chain as a thread embedded in this medium. The expression thus obtained is
\begin{equation}
\label{EMAG}
	G\approx0.82k_{B}T\rho^{4/3}L^{-1/3}_{p}.
\end{equation}
Comparison of the above scaling predictions raises the question of which theoretical approach (if either) better describes the viscoelastic behaviour of semi-flexible polymer solutions in the tightly entangled concentration regime. Existing experimental measurements of the concentration dependence of the plateau modulus \cite{Morse1,Hinner}, are not sufficient to answer that question, because the two putative values of the exponents are numerically quite close ($G\propto\rho^{1.4}$ vs. $G\propto\rho^{1.3\bar{3}}$), so that they both fit the experimental data with reasonable accuracy \cite{Morse1}.

To test the scaling predictions, we experimentally analysed solutions of actin filaments (F-actin), a semi-flexible polymer derived from muscle tissue. As we shall show, F-actin has the useful property that its persistence length can be controlled by varying only the ionic properties of its solvent, without altering other system parameters such as solvent viscosity, polymer concentration or molecular weight, thus allowing us to discriminate between the EMA and BCA models. Hence, this is a rare example of biology helping to answer questions of interest to physics, rather than {\em vice versa}.

At low ionic strength {\em in vitro}, actin exists in the monomeric (globular) G-actin form. G-actin is roughly spherical with a diameter of about 5 nm. When the ionic strength of a G-actin solution is increased to a physiological value (0.1 M), G-actin self-associates, to form F-actin, which is characterized by a persistence length of \mbox{2--20 $\mu$m} (depending on the buffer used) \cite{Isambert2, Gittes} and a diameter of approximately \mbox{8 nm} \cite{Isambert2, MacKintosh}. In order to produce actin filaments with different mechanical properties ($L_{p}$), two sets of F-actin solutions (named System 1 and 2) were prepared using two different buffer recipes, which will be referred to hereafter as \emph{F-buffer 1} and \emph{F-buffer 2}. In both the cases, the initial solutions of G-actin were prepared with the same buffer recipe (\emph{G-buffer}). Polymerization was initialized by increasing the ionic strength of the \emph{G-buffer} to that of the \emph{F-buffers}. The buffer recipes are as follows:
\emph{G-buffer}: 0.2 mM ATP, 0.2 mM CaCl2, 2 mM Tris-HCl, 0.5 mM DTT, pH 8.0.
\emph{F-buffer 1}: 50 mM KCl, 2.0 mM free MgCl2, 5 mM Tris-HCl, 1 mM ATP buffer pH 7.5.
\emph{F-buffer 2}: 25 mM KCl, 1.0 mM free MgCl2, 1.0 mM EGTA, 10 mM MOPS buffer pH 7.0.

In order to obtain independent measurements of the complex modulus $G^{*}$ and of $L_{p}$, the two sets of F-actin solutions were each investigated by two different techniques: passive video particle tracking microrheology (PVPTM) and dynamic light scattering (DLS). 

\begin{figure}
	\centering
		\resizebox{80mm}{!}{\includegraphics{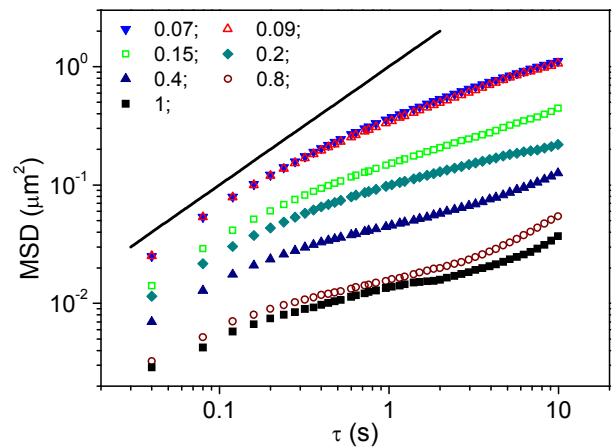}}
	\caption{\label{MSD}Mean-square displacement (MSD) {\em versus} lag-time, for \mbox{0.489 $\mu$m}--diameter beads in F-actin solutions at different concentrations [mg/mL] (System 2). The solid line represents the MSD of the beads suspended in water at 25$^{o}$C.}
\end{figure}

\begin{figure}
	\centering
		\resizebox{80mm}{!}{\includegraphics{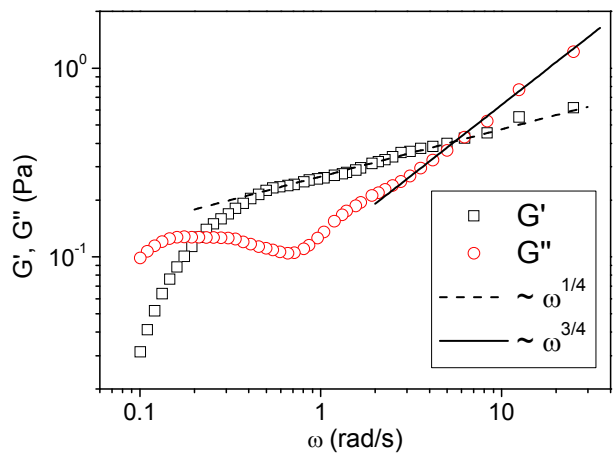}}
	\caption{\label{GpGpp}Storage and loss moduli {\em versus} frequency, for F-actin solution at a concentration 1 mg/mL (System 2). The lines are guides for the eye. The same features are also exhibited by System 1}
\end{figure}

\begin{figure}
	\centering
		\resizebox{80mm}{!}{\includegraphics{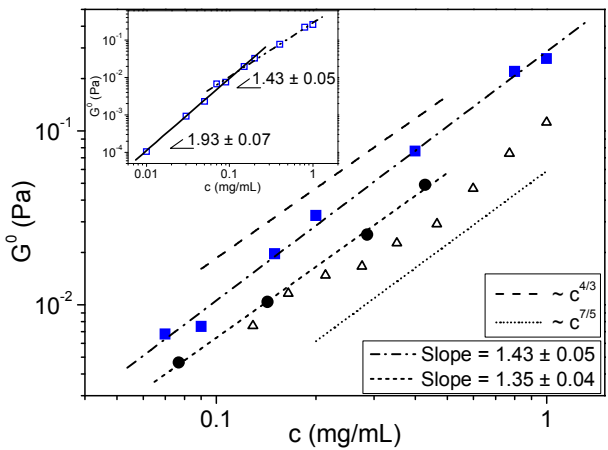}}
	\caption{\label{PlateauM}Plateau modulus $G^{0}$ {\em versus} F-actin concentration, for Systems 1 (circle) and 2 (square). The open triangles are bulk rheology results taken from ref. \cite{Hinner}. The short dash and dash--dot lines are linear fits to the data on the log-log plots (power laws); the dot and dash lines show the BCA ($G^{0}\propto c^{7/5}$) and EMA ($G^{0}\propto c^{4/3}$) scaling predictions respectively. The inset (System 2) shows the actual existence of a different power-law regime at concentrations lower than the \emph{tightly entangled regime} (c.f. Ref. \cite{Morse5}).}
\end{figure}

The first technique, PVPTM, exploits the relationship between the viscoelastic properties of a fluid under investigation, and the mean-square displacement (MSD) of probe particles (of radius $a$), suspended in the fluid and executing Brownian motion (Fig.~\ref{MSD}). This relationship is given by the Generalized Stokes-Einstein relation,
\begin{equation}
\label{GSE}
	\tilde{G}(s)=k_{B}T/\pi as\langle \Delta \tilde{r}^{2}(s)\rangle,
\end{equation}
where $\langle \Delta \tilde{r}^{2}(s)\rangle$ and $\tilde{G}(s)$ are the Laplace-transforms of, respectively, the mean square displacement and the time derivative of the shear modulus. Moving from Laplace space to Fourier space is accomplished by substituting the Laplace frequency $s$ with $i\omega$, so that the real and imaginary parts of $\tilde{G}(i\omega)$ correspond to the storage  and loss  moduli, $G'(\omega)$ and $G''(\omega)$ respectively, with $\omega$ the Fourier frequency (Fig.~\ref{GpGpp}). As probe particles, we used  carboxylate-coated polystyrene beads of diameter \mbox{0.489 $\mu$m}.
It is well known that microrheology can fail to emulate macroscopic results if the probe particle's size and surface chemistry are incorrectly chosen. Microrheological measurements of viscoelastic moduli are often found to match the scaling laws found by macroscopic rheology, whilst disagreeing by a constant factor in absolute magnitude. In the present case, we only require reliable measurements of scaling exponents, not of absolute values. Nevertheless, Fig.~\ref{PlateauM} demonstrates that the absolute values of our measurements are at least as accurate as macroscopic rheometric data in the literature, so that we can be confident of their validity.

In all our PVPTM measurements, we found very good agreement with Morse's scaling predictions, for both $G'(\omega,c)$ and $G''(\omega,c)$ as functions of frequency and concentration. Figure \ref{PlateauM} shows the plateau modulus $G^{0}$ (determined as the value of the storage modulus $G'(\omega)$ at the frequency for which the ratio $G''(\omega)/G'(\omega)$ is minimum) {\em versus} F-actin concentration, for Systems 1 and 2 in the tightly entangled regime ($\approx$ 0.1--2 mg/mL).
It is clear from the figure that, as previously found \cite{Morse1}, the concentration dependence of $G^{0}$ is not able to discriminate between the two predictions (EMA and BCA). However, using the microrheology data, the two theories yield very different predictions for the ratio of persistence lengths in the two systems:
\begin{equation}
\label{BCAratio}
	BCA\Longrightarrow L_{p1}/L_{p2} 
	= \left(G^{0}_{2}/G^{0}_{1}\right)^5 = 18\pm3
\end{equation}
\begin{equation}
\label{EMAratio}
	EMA\Longrightarrow L_{p1}/L_{p2}
	= \left(G^{0}_{2}/G^{0}_{1}\right)^3 = 6.2\pm0.7
\end{equation}
Hence, we can distinguish between the two models by measuring the actual persistence lengths. For that purpose, DLS measurements were performed on some of the samples. 

\begin{figure}
	\centering
		\resizebox{80mm}{!}{\includegraphics{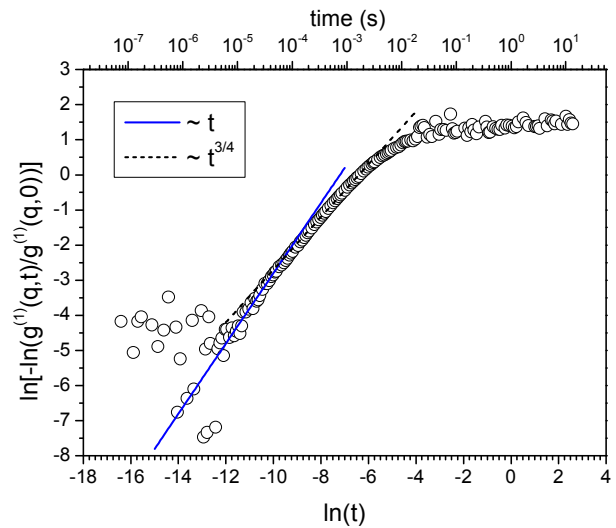}}
	\caption{\label{CF}Double logarithm of the normalized dynamic structure factor $g^{(1)}(q,t)/g^{(1)}(q,0)$ {\em versus} logarithm of time, at a scattering angle of $90^o$, for a solution at F-actin concentration \mbox{$c = 0.4$ mg/mL} (System-1). Similar data are obtained for System 2. The lines are guides for the eye.}
\end{figure}

Kroy and Frey \cite{KroyFrey} provide a simple analytical expression for the time-dependent decay of the dynamic structure factor $g^{(1)}(q,t)$ of semi-flexible polymers in semi-dilute solution, where $q=[(4\pi n)/\lambda_{0}]\sin(\theta/2)$ is the magnitude of the scattering wave-vector, defined as the difference between the incident and scattered wave-vectors, $n$ is the refractive index of the solvent, $\lambda_{0}$ is the wavelength of the laser {\em in vacuo}, and $\theta$ is the scattering angle. 
They showed that, at sufficiently long times ($t>>(qL_{p})^{-4/3}\gamma^{-1}_{q}$), the dynamic structure factor reduces to a simple stretched exponential,
\begin{equation}
\label{stretchedexp}
	g^{(1)}(q,t)=g^{(1)}(q,0)\exp\left[-\Gamma(1/4)(\gamma_{q}t)^{3/4}/3\pi\right],
\end{equation}
where $\Gamma(1/4)=3.62561$ and decay rate $\gamma_{q}$ given by
\begin{equation}
\label{gammas}
	\gamma_{q} = 
	k_{B}Tq^{8/3}\left[5/6-\ln(qa_{h})\right]/4\pi\eta_{s}L^{1/3}_{p},
\end{equation}
where $\eta_{s}$ is the solvent viscosity. They also calculated the form of the dynamic structure factor in the limit $t\to0$, which can be used to measure the microscopic hydrodynamic lateral diameter $a_{h}$ of the polymer chain:
\begin{equation}
\label{singleexp}
	g^{(1)}(q,t)=g^{(1)}(q,0)\exp\left(-\gamma^{0}t\right)
\end{equation}
with initial decay rate
\begin{equation}
\label{gamma0}
	\gamma^0=k_{B}Tq^{3}\left[5/6-\ln(qa_{h})\right]/6\pi^{2}\eta_{s}.
\end{equation}
The validity of Eqs.~(\ref{stretchedexp}--\ref{singleexp}) requires \cite{KroyFrey} the conditions $a_{h}<<q^{-1}<<L_{p}\sim L$ and $q^{-1}<<L_{m}$, all of which are satisfied by all of our samples, since \mbox{$a_{h}\approx 9$ nm}, \mbox{$q^{-1}\approx 125-33$ nm} (where only the high-$q$ data will be used in the final result), while $L_{p}$ and $L$ are of order of microns and $L_{m}\approx0.51-0.16\mu$m \cite{Käs}.
It is clear from Fig.~\ref{CF} that Kroy and Frey's calculations \cite{KroyFrey} are applicable to our experimental system, despite the scatter of data at early times. Indeed, $g^{(1)}(q,t)$ decays initially as a simple exponential (identified by the line with unit slope in Fig.~\ref{CF}) and exhibits stretched exponential behavior at longer times (line with slope $3/4$).

\begin{figure}
	\centering
		\resizebox{80mm}{!}{\includegraphics{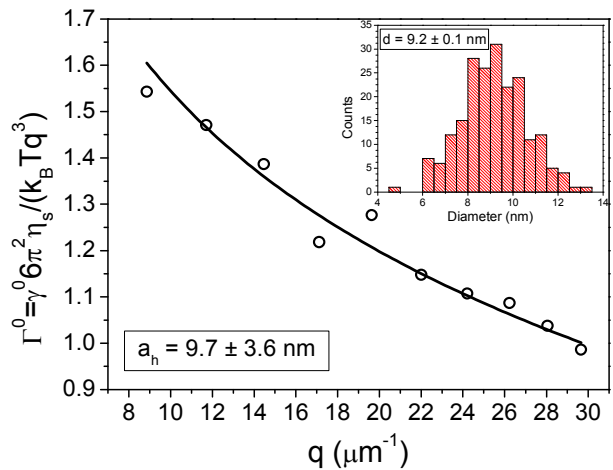}}
	\caption{\label{Diameter}Normalized initial decay rate $\Gamma^{0}$ {\em versus} scattering wave-vector $q$, and the best fit of Eq.~(\ref{gamma0}) using $a_{h}$ as a free parameter, for a solution at F-actin concentration \mbox{$c = 0.1$ mg/mL} (System-2). Insert: Diameter distribution from TEM image analysis.}
\end{figure}

The hydrodynamic diameter $a_{h}$ was obtained from the data by evaluating the initial decay rate, $\gamma^0$, via a linear fit to the logarithm of the normalized dynamic structure factor, in a time window between \mbox{$10^{-7}$ s} and \mbox{$4\times10^{-5}$ s}. From Eq.~(\ref{gamma0}) the non-dimensionalized initial decay rate, $\Gamma^{0}=6\pi^{2}\eta_{s}\,\gamma^0/(k_{B}Tq^{3})$, was then fitted by $[5/6-\ln(qa_{h})]$ (Fig.~\ref{Diameter}) to obtain $a_{h}$.
For all the F-actin solutions investigated (as with other large polymers \cite{Carrick}), we found very good agreement between the indirect measurement of the lateral hydrodynamic diameter (averaged value \mbox{$\overline{a_h}=9\pm3$ nm}) and that measured directly by transmission electron microscopy (TEM) image analysis (\mbox{$\overline{a_h}=9.2\pm0.1$ nm}; see Fig.~\ref{Diameter} insert).

\begin{figure}
	\centering
		\resizebox{80mm}{!}{\includegraphics{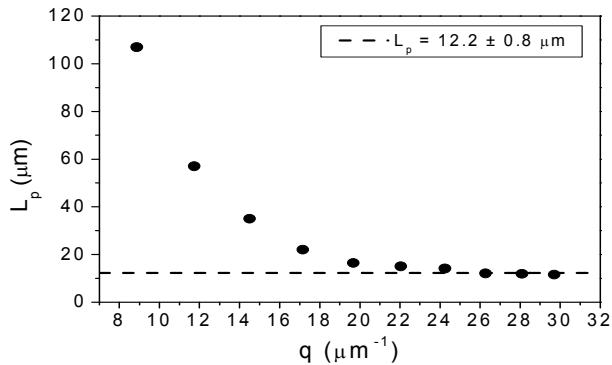}}
	\caption{\label{Lp}Persistence length $L_{p}$ {\em versus} wave-vector $q$ for a solution at F-actin concentration \mbox{$c = 0.2$ mg/mL} (System-1) (derived using \mbox{$a_{h}=9.2$ nm}).}
\end{figure}
With the hydrodynamic diameter now determined, it can be used in Eqs.~(\ref{stretchedexp}) and (\ref{gammas}) to estimate the persistence length $L_{p}$. We adopt an average value \mbox{$a_h=9.2$ nm} for both systems. Equations~(\ref{stretchedexp}) and (\ref{gammas}) were fitted to the measured time-dependence of the dynamic structure factor, for various scattering vectors $q$, with $L_p$ as the only fitting parameter. The results for System 1 are shown in Fig.~\ref{Lp}. One might expect the result to be independent of $q$ but, as noted in Ref.~\cite{KroyFrey}, the equations apply only to scattering vectors large enough to resolve single filaments. For scattering vectors smaller than the inverse mesh size $L_{m}$, the structure factor is averaged over several filaments and Eqs.~(\ref{stretchedexp}) and (\ref{gammas}) no longer hold. It is therefore the large-$q$ asymptote that marks the true persistence length in Fig.~\ref{Lp}. We thus consider only averaged values of $L_{p}$ for \mbox{$q\geq22$ $\mu$m$^{-1}$}, as reported for the two systems in Table \ref{tab:Lp}, where three independent measurement of the ratio of persistence lengths are all in agreement.

\begin{table}[b]
	\medskip
	\begin{tabular}{|c|c|c|c|}
	\hline
	Actin concentration & $L_{p1}$  & $L_{p2}$  & $L_{p1}/L_{p2}$ \\
	(mg/mL) & ($\mu$m) & ($\mu$m) & \\
	 \hline\hline
		0.1 & $10\pm1$ & $1.88\pm0.05$ & $5.3\pm0.6$ \\
		0.2 & $12.2\pm0.8$ & $2.24\pm0.15$ & $5.4\pm0.5$ \\
		0.4 & $18\pm2$ & $3.16\pm0.09$ & $5.7\pm0.7$ \\
		\hline
	\end{tabular}
	\caption{\label{tab:Lp}Persistence lengths in Systems 1 and 2. We also found a weak concentration dependence of $L_{p} \propto c^{1/3\pm0.04}$ for both Systems. This functionality is consistent with the definition of $L_{p}$, which is expected to grow from $L_{p}=\kappa/k_{B}T$, in dilute solution (where $\kappa$ is the bending modulus), up to $L_{p}\equiv\infty$, in the nematic phase.}
\end{table}

In conclusion, the true ratio of persistence lengths in the two systems is consistent with the value predicted by Morse's EMA approximation (equation \ref{EMAG}) \cite{Morse1}, but is more than four standard deviations away from that predicted by the more established BCA scaling law (equation \ref{BCAG}).
We have thus resolved the controversy, and established that the effective medium approximation more accurately models the tightly-entangled regime of semi-flexible polymer solutions. The strength of this result rests on the clear agreement between two completely independent experimental techniques.

ACKNOWLEDGMENTS: We thank Peter Olmsted and Tanniemola Liverpool for helpful conversations. The work was funded by the EPSRC, and RMLE is funded by the Royal Society.

\end{document}